\documentstyle[aps,twocolumn,prl,epsf]{revtex}
\begin{document}

\def\Cml{Coupled map lattice}
\def\Cmls{Coupled map lattices}
\def\cml{coupled map lattice}
\def\cmls{coupled map lattices}
\def\e{\varepsilon}
\def\ds{\displaystyle}
\def\PTP{Prog.~Theor.~Phys.}
\def\PLA{Phys.~Lett.~A}
\def\PL{Phys.~Lett.}
\def\PRL{Phys.~Rev.~Lett.}
\def\PRE{Phys.~Rev.~E}
\def\PD{Physica D}
\def\JSP{J.~Stat.~Phys.}
\def\IJBC{Int.~J.~Bifurcation and Chaos}
\def\JTB{J.~theo.~Biol.}
\def\JAE{Journal of Animal Ecology}

\input amssym.def
\input amssym.tex

\draft

\title{\bf Low dimensional travelling interfaces in
           coupled map lattices}

\author{R.\ Carretero-Gonz\'alez\thanks{e-mail: R.Carretero@ucl.ac.uk}
        \thanks{Current address: Center for Nonlinear Dynamics and its
        Applications (CNDA), Dept.~Civil and Environmental Engineering,
        University College London, Gower Street, London, WC1E 6BT},}

\address{
        School of Mathematical Sciences,
        Queen Mary and Westfield College,
        Mile End Road, London E1 4NS, U.K.}

\date{To appear in {\it Int.~J.~Bif.~Chaos}}

\maketitle

\noindent {\bf Keywords:}
Coupled map lattices, travelling waves, mode-locking, circle map, rotation
number, bifurcation.

\begin{abstract}
We study the dynamics of the travelling interface arising from a bistable 
piece-wise linear one-way coupled map lattice. We show how the dynamics of the
interfacial sites, separating the two superstable phases of the local map,
is finite dimensional and equivalent to a toral map. The velocity of the
travelling interface corresponds to the rotation vector of the toral map. As
a consequence, a rational velocity of the travelling interface is subject to
mode-locking with respect to the system parameters. We analytically compute
the Arnold's tongues where particular spatio-temporal periodic orbits exist.
The boundaries of the mode-locked regions correspond to border-collision
bifurcations of the toral map. By varying the system parameters it is
possible to increase the number of interfacial sites corresponding to a
border-collision bifurcation of the interfacial attracting cycle. We finally
give some generalizations towards smooth coupled map lattices whose interface 
dynamics is typically infinite dimensional.
\end{abstract}

\section[Introduction]{Introduction}

{\Cmls} (CML) where introduced by Kaneko [1983, 1984] as a paradigm for the
study of spatio-temporal complexity such as turbulence [Kaneko, 1986, 1989,
Beck 1994], convection [Yanagita \& Kaneko, 1993], open flows [Willeboordse
\& Kaneko 1995, Kaneko 1985], patch population dynamics [Hassell {\it et
al.}, 1995, Sole \& Bascompte 1995], etc. An interesting spatio-temporal
feature is the emergence of coherent travelling structures [Kaneko 1992,
1993] along the lattice. In this paper we use a simple example of a
piece-wise linear one-way CML with two superstable basins in order to
restrict the front dynamics to a finite dimensional system. At the beginning
of Sec.~2 we recall some features in a particular region of the parameter
space where the travelling interface consist of a single site. The dynamics
in this case can be reduced to a one-dimensional map of the circle. The
velocity of the travelling interface is computed by means of the rotation
number of the circle map and thus the mode-locking regions of the velocity
with respect to the system parameter are computed analytically. Later in
Sec.~2 we perform an equivalent approach for the case of a 2 sites
travelling interface. Here the dynamics of the travelling interface can be
reduced to a two-dimensional toral map whose rotation vector gives the
travelling velocity. We give a method for computing the mode-locking regions
in this case. In Sec.~3 we present the Arnold's tongues for higher
dimensional interfaces and show the bifurcation diagram as the number of
sites in the interface is increased. Finally we give some comments on the
generality of our results for generic bistable local maps whose attractors
do not possess superstable basins and thus their corresponding travelling
interface is typically infinite dimensional.

A CML is a dynamical system with discrete space, discrete time and {\sl
continuous} state space. We could think of CMLs as a generalization of
cellular automata whose state space is discrete [Chate \& Manneville, 1989,
1990]. Consider a one-dimensional collection of cells labeled by the integer
index $i$ and with dynamical variable $x_t(i)$ at time $t$. The lattice
could be infinite or finite with periodic or fixed boundary conditions. For
the present study the particular choice of lattice is not crucial since we
deal with the propagation of finite dimensional localized structures with
spatio-temporal periodicity. Therefore, for the numerical experiments, it
suffices that the lattice is large enough to enclose the localized front for
a whole period. The CML dynamics consists of two independent stages: the
local dynamics and the coupling dynamics. The former is the application of
the one-dimensional {\it local map\/} $f_i(x)$ to every site and the latter
couples the dynamics by means of a weighted sum over a neighbourhood
$\nu_i$:
\begin{equation}
\begin{array}{rcl}
x_t(i) &\overrightarrow{\quad\rm{local}\quad} & f_i(x_t(i)) \\[3.0ex]
       &\overrightarrow{\quad\rm{coupling}\quad} &
x_{t+1}(i)=\sum_{\nu_i}{\e_{\nu_i} f_{\nu_i}(x_t(i))},
\end{array}
\end{equation}
where the coefficients $\e_{\nu_i}$ determine the type of coupling
interaction. A physically meaningful interaction will have typically a
limited range, with decreasing strength for distant neighbours. The local
map and neighbourhood are often taken to be the same for all sites ($f_i=f$
and $\nu_i=\nu$) indicating an homogeneous dynamics ---{\it homogeneous\/}
CML. Also one asks that $\sum{e_{\nu_i}}=1$ as a conservation law, since
failure to do so may lead to non-boundedness of the state as time tends to
infinity.

The two most widespread models of CML are
\begin{equation} \label{diffu}
x_{t+1}(i)= (1-\e)f(x_t(i))+{\e\over 2}\left( f(x_t(i-1))+f(x_t(i+1))\right)
\end{equation}
and
\begin{equation} \label{one-way}
x_{t+1}(i)= (1-\e) f(x_t(i)) + \e f(x_t(i-1)),
\end{equation}
which are called {\it diffusive\/} and {\it one-way\/} CML respectively.
The coupling parameter $\e$ is set to satisfy $0\leq\e\leq 1$.

In this paper we study interfaces separating two different phases in
bistable one-way CMLs. For two different phases to coexist in the lattice we
consider a local map with two and only two stable fixed points $x_-^*$ and
$x_+^*$ ($x_-^*<x_+^*$), {\it i.e.}~a bistable local map, then each of the
two homogeneous states $x_t(i)=x^*_\pm$ is stable for the CML dynamics. We
now consider {\it minimal mass\/} states of the lattice whose sites are
arranged in non-decreasing order ($x_t(i+1)\geq x_t(i)$), and choose local
maps that are continuous and non-decreasing. For such maps it is possible to
have an interface, sites separating the stable $x^*_\pm$ phases, that
travels along the lattice [Carretero {\it et al.}~1997a, 1997b]. Travelling
fronts in generic bistable CMLs are typically infinite dimensional since the
attraction towards the stable points $x^*_\pm$ is an infinite process.
Nevertheless the convergence towards $x^*_\pm$ is exponential (for stable
fixed points). This exponential convergence ensures a localization of the
front despite the fact that the interface involves an infinite number of
sites.

The simplest of all the local maps with two superstable regimes is the
family of piece-wise linear local maps $f_a$ defined by
\begin{equation} \label{fa} f_a(x)=\left\{ \begin{array}{cll}
                                       -1&\mbox{if}& x\leq -a\\[2.0ex]
{\ds 1\over \ds a}\,x&\mbox{if}& -a<x<a\\[2.0ex]
                                       1 &\mbox{if}& x\geq a
\end{array}\right.\qquad 0 < a < 1,\end{equation}
and by the step function between $-1$ and $1$ centered at the origin when
$a=0$. The fixed points $x_-^*=-1$ and $x_+^*=1$ are superstable, with
superstable basins $S_-=[-1,-a]$ and $S_+=[a,1]$ respectively, and thus the
homogeneous phases $x(i)=x_\pm^*$ are superstable as well. The parameter
space is now $(\e,a)\in[0,1]\times[0,1]$. The piece-wise linear family $f_a$
possesses the important property of collapsing orbits that get $(1-a)$-close
to $x^*_\pm$. The propagating front is then subject to a cut-off near
$x^*_\pm$ ensuring a finite number of interfacial sites (falling in the
linear regime $[-a,a]$, for $0<a<1$). Therefore, instead of dealing with an
infinite dimensional interface ---that is the case for generic CMLs--- we
are let with a finite number of interfacial sites allowing us a precise
description of the front dynamics.

\section[One- and Two-Dimensional Interface Dynamics]
        {One- and Two-Dimensional Interface Dynamics}

\subsection[One-dimensional interface dynamics]
           {One-dimensional interface dynamics}

In a previous paper [Carretero {\it et al.}, 1997a] we showed how the
dynamics of the whole lattice could be reduced to a one-dimensional map for
a particular region of the parameter space. We now briefly recall some of
the results in order to use the same approach for the $N$-dimensional case.

Consider a general minimal mass state with $N$ sites in the interface. The
interface, separating the two phases, is well-defined in the case of the
piece-wise linear map $f_a$ since any site falling into the superstable
regions $S_\pm$ is mapped to $\pm 1$ by $f_a$ in one iteration. Therefore we
define the {\it interface\/} as the collection of sites belonging to the
unstable region $U=[-a,a]$. We will often use the term interface for
denoting the interval $U$ itself. The general form of a minimal mass state
is
$$X_t=(\,\ldots, -1,-1,x_t(k+1),\ldots,x_t(k+N),1,1,\ldots\,),$$
with $|x_t(i)|<a$ for $i=k+1,\ldots,k+N$. Without loss of generality one may
choose $k=0$. The image of $X_t$ by the one-way CML (\ref{one-way}) is
\begin{equation}\label{X_(t+1)}
\begin{array}{rcl}
X_{t+1}&=& (\ldots,-1,g(-a,x_1),g(x_1,x_2),\dots,\\[1.0ex]
       & & g(x_{N-1},x_N),g(x_N,a),1,\ldots\,),
\end{array}
\end{equation}
where now the explicit time dependence is dropped and the subscript denotes
the label of the site. The function $g$, defined by
\begin{equation} \label{g}
   g(x_1,x_2)={\ds 1-\e\over\ds a}\,x_2+{\ds\e\over\ds a}\,x_1,
\end{equation}
originates from the combination of the local dynamics and the coupling. At
both extremes of the interface the function $g$ only uses explicitly one
site since the other is $-1$ or $1$, left and right extremes respectively.
By defining now the maps
\begin{equation} \label{fPlusMinus} \begin{array}{rcccl}
   f_0(x)&=&g(-a,x)&=&{\ds 1-\e\over \ds a}\,x-\e\\[2.0ex]
   f_1(x)&=&g(x,a) &=&{\ds \e\over \ds a}\,x + (1-\e),
\end{array} \end{equation}
it is possible to write (\ref{X_(t+1)}) as
\begin{equation}\label{X_(t+1)b}
\begin{array}{rcl}
X_{t+1}&=&(\ldots,-1,f_0(x_1),g(x_1,x_2),\dots,\\[1.0ex]
       & &g(x_{N-1},x_N),f_1(x_N),1,\ldots\,),
\end{array}
\end{equation}
where now the map $f_0$ ($f_1$) originates from the combination of the local
map and the interaction between the left-most (right-most) site of the
interface with the neighbouring homogeneous region, while the map $g$
originates from the interaction between interfacial sites.

By defining
\begin{equation}\label{GapSize}
\gamma(\e,a)\,=\,\gamma_--\gamma_+
\,=\,{\frac{\ds a(1-a-2\e(1-\e))}{\ds\e(1-\e)}},
\end{equation}
where
$$
\gamma_-={a(\e - a)\over 1-\e}\qquad \gamma_+={a(a+\e-1)\over\e},
$$
it is possible to show [Carretero {\it et al.}, 1997a] that if $\gamma>0$
the state of the whole lattice is reduced to the one-dimensional circle map
\begin{equation} \label{Phi}
\Phi_{\e,a}(x)=\left\{ \begin{array}{cll}
        f_1(x)&\quad\mbox{if }& x \in [-a,\gamma_+]\\[2.0ex]
          a   &\quad\mbox{if }& x \in \Gamma\\[2.0ex]
        f_0(x)&\quad\mbox{if }& x \in [\gamma_-,a]
\end{array}\right.,
\end{equation}
where $\Gamma=(\gamma_+,\gamma_-)$ and its length (with sign) is given by
(\ref{GapSize}) and is referred as the {\it gap\/}. The 2-parameter map of
the circle $\Phi_{\e,a}$, called {\it auxiliary map\/}, accounts for the
dynamics of the only interfacial site in the non-negative gap case. The
auxiliary map maps the interfacial site throughout the evolution of the
interface and the position of the latter is given by a symbolic dynamics
representation.

\begin{figure}
\centerline{\epsfxsize=8.5cm \epsfbox{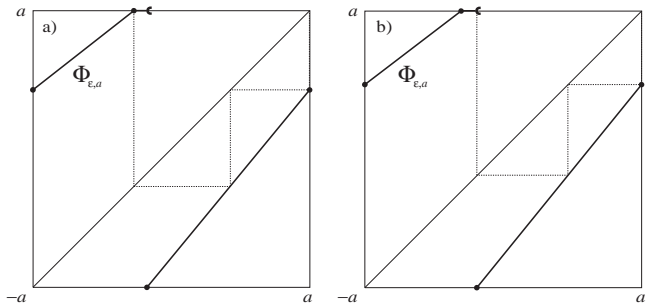}}
\vskip .25cm
\caption{Moments of birth a) and death b) of
the $v=1/3$ tongue for $a=0.5$. The $v=1/3$ tongue begins (ends) when the
interfacial orbit, mapped through the auxiliary map $\Phi_{\e,a}$, collides
with the left-most (right-most) extreme of the gap.}
\label{border.eps}\end{figure}

The auxiliary map is a map of the circle and it is possible to show
[Carretero {\it et al.}, 1997a] that the velocity of the travelling
interface is given by its rotation number. The superstable region (the gap)
of the auxiliary map induces a mode-locking of the rotation number as the
parameters are varied. All the mode-locking regions, Arnold's tongues, where
any particular rational velocity $0\leq v(\e)=p/q\leq 1$ exists, may be
computed analytically by finding the $(\e,a)$-interval for which the gap
interval intersects the identity line after $q$ iterations passing $p$ times
through the upper region of the auxiliary map. The extremes cases for this
to happen arise from a {\it border-collision bifurcation} [Maistrenko {\it
et al.}, 1995] when an orbit starting in the gap falls at the beginning
(end) of the gap interval after $q$ iterations. As an example we show in
Fig.~\ref{border.eps} the $v=1/3$ plateau limit orbits for $a=0.5$.

\subsection[Two-dimensional interface dynamics]
           {Two-dimensional interface dynamics}

When the gap size is non-negative the whole lattice was reduced to the
one-dimensional auxiliary circle map. For negative gap size there are more
than one site in the interface at the same time. Thus a one-dimensional
reduction of the whole lattice does not seem attainable. However, thanks to
the superstable basins $S_\pm$, the number of sites in the interface is
still finite ---less than or equal to $N$--- and one could reduce the
dynamics of the whole lattice to a $N$-dimensional system containing the
information of the interface sites.

\subsubsection[Minimal $N$-state layers]{Minimal $N$-state layers}

Before reducing the dynamics, one has to find the parameter regions where a
particular number of sites is present in the interface. Define a {\it
minimal mass $N$-state} or {\it minimal $N$-state} to be a minimal mass
state having, during its evolution, a minimum of $N-1$ sites in the
interface. The spatial discretization induces a change in the number of
sites in the interface, so that the number of sites in the interface for a
minimal $N$-state cannot be constant, it varies during its evolution between
$N-1$ and $N$.

Suppose that the front shape at any time $t$ is given by $x_t(i)=h(i-vt)$,
where $v$ is its travelling velocity and $h$ is a real valued function
independent of $t$. The sites falling in the interface, are the sites
contained in the interval $H=(h^{-1}(-a),h^{-1}(+a))$. Since the length of
$H$ and the distance between nodes remains constant, the number of sites in
$H$ can only vary by one during the evolution. In the case of a minimal
$N$-state the number of sites in the interface is then $N$ or $N-1$.
Nevertheless it is possible to remove this duality by a simple trick as it
is shown in the next section. Using the same notation, the non-negative gap
case corresponds to a minimal 1-state.

In order to show that a minimal $N$-state possesses $N$ or $N-1$ sites in
the interface we supposed that the front shape is given by a function $h$.
If the interfacial orbit is periodic it is easy to construct the front shape
$h$ by considering any curve $h(i-vt)$ passing through all the points of the
orbit. It is interesting to notice that such $h$ is not unique, in fact
there exists an infinite number of possible $h$-functions passing through
the finite collection of required points. On the other hand, when the
velocity is irrational the interfacial orbit consists of an infinite number
of distinct points. It is possible to reconstruct the front shape for
irrational $v$ by superimposing snapshots of the front shape taken in a
co-moving reference frame with velocity $v$. Extensive numerical
experimentation suggests that the travelling interface is indeed given by an
increasing function. Further reconstructions of travelling interfaces in
more general types of CMLs are consistent with this result [Carretero {\it
et al.}, 1997c].

\begin{figure}
\centerline{\epsfxsize=8.5cm \epsfbox{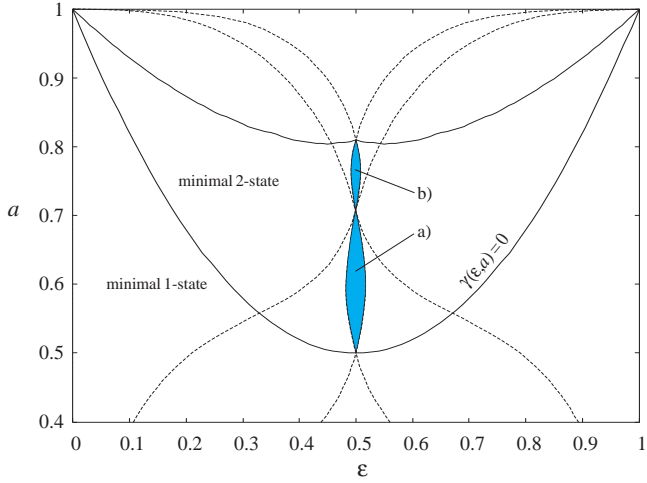}}
\vskip .25cm
\caption{The $v=1/2$ tongues in the minimal
2-state layer. The tongues a) and b) are obtained by solving the
inequalities (\ref{cond1-2}) corresponding to the two possible
period-2 orbits.}
\label{blobs1-2.eps}\end{figure}

\begin{figure}
\centerline{\epsfxsize=8.5cm \epsfbox{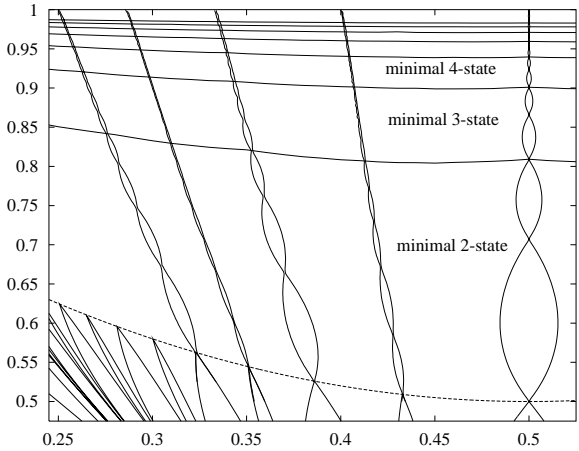}}
\vskip .25cm
\caption{Tongues in the $N$-dimensional minimal
$N$-state layers. The tongues for the velocities $v=1/2,\,2/5,\,1/3,\,2/7$
and $1/4$  (from right to left) in the minimal $N$-state layers for
$N=1,\dots,8$ are presented.}
\label{blobs-z.ps}\end{figure}

In Fig.~\ref{blobs1-2.eps} we depict the $(\e,a)$ region where the minimal
2-state is present. The boundary between the minimal 1-state and the minimal
2-state layers is the zero-gap curve. Above that curve, the gap is negative,
and there appears to be an infinite family of layers, labeled by $N$, and
separated by boundary curves (cf.~Fig.~\ref{blobs-z.ps}).

\subsubsection[Two-dimensional auxiliary map]{Two-dimensional auxiliary map}
Throughout this section consider the case $N=2$, {\it i.e.}~take
$(\e,a)$-parameter values lying in the minimal 2-state layer. One would like
to derive an auxiliary map that contains all the information of the
interface dynamics as the auxiliary map $\Phi_{\e,a}$ does for the
non-negative gap case. In order to find such map one has to take into
account all possible evolution combinations of a minimal 2-state.
The general form for a minimal 2-state $X_t$ is
$$X_t \,=\,(-1,\,x_1,\,x_2,\,1,\dots),$$
but because the size of the interface may be 1 or 2 one has the two
possibilities $-a<x_1<x_2<a$ or $-a<x_1<a\leq x_2\leq 1$. Therefore, the
first site, $x_1$, is always in the interface while the second, $x_2$, may
or may not be in the interface. If the second site is not in the interface
one could assign to it any value between $+a$ and $+1$ and its image by the
piece-wise linear map $f_a$ would remain the same. In this case we choose to
reduce $a\leq x_2\leq 1$ to $x_2=a$ and the dynamics is not altered. Using
this reduction the site $x_2$ is now always included in the interface and it
would be possible to remove the duality of having $2$ or $1$ sites in the
minimal $2$-state layer.

\begin{figure}
\centerline{\epsfxsize=8.5cm \epsfbox{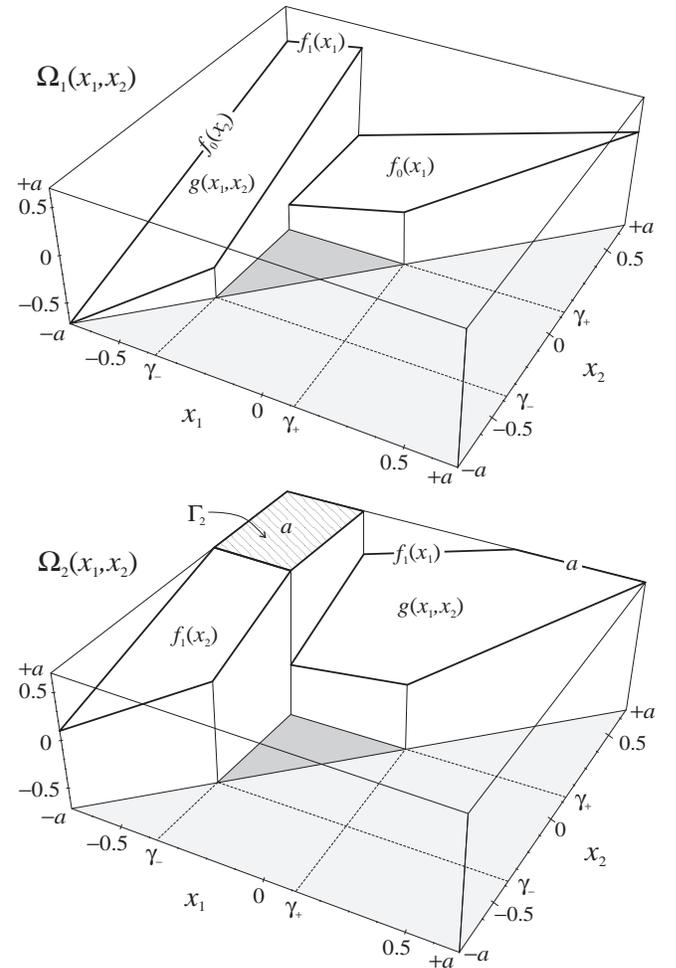}}
\vskip -.25cm
\caption{Example of the two-dimensional auxi\-lia\-ry
map $\Omega(x_1,x_2)$$=$$\left(\Omega_1(x_1,x_2),\Omega_2(x_1,x_2)\right)$
accounting for the in\-ter\-fa\-cial sites dynamics in the minimal 2-state for
$(\e,a)=(0.365,0.7)$.}
\label{omegas.eps}\end{figure}

It is possible to prove that the pair of interfacial sites $(x_1,x_2)$ is
mapped to $(x'_1,x'_2)$ in 6 different ways depending on their value. The
dynamics then induces a two-dimensional map $\Omega(x_1,x_2)=(x'_1,x'_2)$.
An example of the map $\Omega$, obtained by considering all the
possibilities is depicted in Fig.~\ref{omegas.eps} by separating it into its
two components $x'_1=\Omega_1(x_1,x_2)$ and $x'_2=\Omega_2(x_1,x_2)$. The
shaded areas in Fig.~\ref{omegas.eps} correspond to unreachable situations
for a minimal 2-state. Since $X_t$ is a minimal mass state, $x_2$ has to be
larger or equal to $x_1$, thus we eliminate the area where $x_2<x_1$
---light shaded area. On the other hand, it is straightforward to prove that 
if $x_1$ and $x_2$ are both at the same time in the interval 
$[\gamma_-,\gamma_+]$, the succesive state is a minimal 3-state and thus we 
eliminate this possibility ---dark shaded area. These forbidden areas are 
never reached by the dynamics.

For the 2-dimensional case (minimal 2-state layer) the auxiliary
map $\Omega$ is a toral map in two dimensions ---a toral map is the
generalization of a circle map to more than one dimension (cf.~[Baesens {\it
et al.}, 1991]). Generalizing the idea of rotation number in more dimensions
one could think of a {\it rotation vector} [Baesens {\it et al.}, 1991]
whose entries correspond to the rotation number in every component. The
rotation vector $\boldmath\rho$ $=(\rho_1,\rho_2)$ of $\Omega(x_1,x_2)$ is
then two-dimensional. But since both sites $(x_1,x_2)$ belong to the same
interface, that is moving with a defined velocity, the two components of the
rotation vector have to be equal ($\rho_1=\rho_2$). In other words, the site
$x_1$ has to travel at the same velocity than the site $x_2$ in order the
interface to remain in the minimal 2-state layer. The only way $x_1$ would
have different rotation number than $x_2$ is for an interface that is
increasing (or decreasing) in size. Therefore instead of taking the whole
rotation vector one may use a single scalar to describe the rotation around
$\Omega(x_1,x_2)$. Thus, the velocity of the travelling interface in the
minimal 2-state layer is given by this scalar, which will be simply called
from now, the rotation number of $\Omega(x_1,x_2)$.

\subsubsection[The tongues for the two-dimensional case]
              {The tongues for the two-dimensional case}
The two-dimensional auxiliary map $\Omega$ is piece-wise linear ---it is a
combination of planes--- and possesses the analogue of the gap $\Gamma$ for
the one-dimensional case: the region $[-a,\gamma_-]\times [\gamma_+,a]$, see
dashed plateau in Fig.~\ref{omegas.eps}. This region, which we refer to as
the {\it two-dimensional gap} and denote by $\Gamma_2$, acts in the same way
as its one-dimensional analogue. Any $(x_1,x_2)$-orbit falling into
$\Gamma_2$ is superstable and therefore is parametrically stable to
perturbations. This parametric stability gives rise to the mode-locking of
the velocity. Thus one expects the minimal 2-state layer to contain Arnold's
tongues in the same way the minimal 1-state layer does.

Let us derive the tongues for $v=1/2$ in the two-dimensional case. First of
all one has to choose the orbit. In the one-dimensional case the orbit was
uniquely determined by a symbolic coding of the velocity for fixed values of
$(\e,a)$ [Carretero {\it et al.}, 1997a]. However, in the two-dimensional
case, there are different possible evolutions for a given velocity. This is
due to the fact that with more sites in the interface a wider selection of
orbits is possible. The two orbits of a minimal 2-state giving a velocity
$v=1/2$ correspond to a) $x_2\geq a$ all the times and b) alternating one 
and two sites in the interface. First of all the orbit has to be periodic
(period two). Therefore we must have a) $f_1(x'_1=f_0(x_1)) = x_1$ and b)
$g(x'_1=f_0(x_1),x'_2=f_1(x_1))=x_1$, {\it i.e.},
\begin{equation}\label{Omega^2(x1)}\begin{array}{lccl}
\hbox{a)}\qquad& x_1&=&{\ds a(a(1-\e)-\e^2)\over\ds a^2-\e(1-\e)},
\qquad\qquad\qquad\qquad\qquad\\[4.0ex]
\hbox{b)}\qquad& x_1&=&{\ds a(1-2\e)\over\ds a^2+2\e(\e-1)}.
\qquad\qquad\qquad\qquad\qquad
\end{array}\end{equation}
Next, one has to verify that the sites fall in the right intervals. That is
\begin{equation}\label{cond1-2}\begin{array}{lccl}
\hbox{a)} \quad& f_1(x_1) > a \hbox{\quad and\quad}
                 f_0(x'_1=f_0(x_1)) < -a,\\[2.0ex]
\hbox{b)} \quad& f_0(x'_1=f_0(x_1)) < -a \hbox{\quad and\quad}
                 f_1(x'_2=f_1(x_1))> a.
\end{array}\end{equation}
Combining Eqs.~(\ref{Omega^2(x1)}) and (\ref{cond1-2}) gives the conditions
that $\e$ and $a$ must satisfy in order to have the period-2 orbits. In
Fig.~\ref{blobs1-2.eps} we display the regions where these conditions are
satisfied in the minimal 2-state layer.

The same procedure may be applied to find the tongues for any rational
velocity in the minimal 2-state layer. For every chosen $0<v=p/q<1$ there
are $q$ possible different orbits, by combining 1 and 2 sites in the
interface, and each one has a corresponding tongue in the minimal 2-state
layer.

\section[$N$-Dimensional Interface Dynamics]
        {$N$-Dimensional Interface Dynamics}

Let us now consider the case of a minimal $N$-state layer for $N>2$. The
dynamics in that layer may be reduced to a $N$-dimensional auxiliary toral
map that controls the $N$-tuplet of sites in the interface. Again, thanks to
the superstable region of the local map, this $N$-dimensional toral map will
have a $N$-dimensional gap $\Gamma_N$ such that any $N$-tuplet in $\Gamma_N$
maps one of its components to $a$. The parametric stability is again induced
by the presence of the gap $\Gamma_N$ when the latter exists.

\begin{figure}
\centerline{\epsfxsize=8.5cm \epsfbox{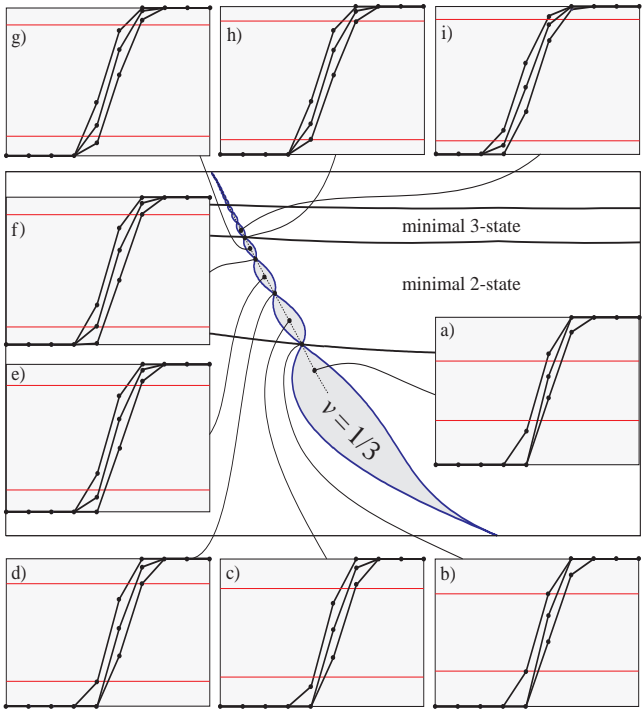}}
\caption{Successive states of the lattice in the
$v=1/3$ tongue. Every time a site touches the boundary of the superstable
region, in order to pass to a new configuration, the tongue $\e$-width is
zero. The states from a) to i) correspond to increasing values of $a$ inside
the $v=1/3$ tongues. The state a) is a minimal 1-state while the state b) is
the transitional state to a minimal 2-state (zero gap). The states c), e)
and g) correspond to the three different possibilities for a minimal 2-state
and d) and f) correspond to the transition points between these three
states. State i) is already is a minimal 3-state and h) is the transition
point between the minimal 2-state to the minimal 3-state.}
\label{bifu1-3.eps}\end{figure}

Thus, in every minimal $N$-state layer there are $q$ tongues corresponding
to the velocity $v=p/q$ given by the $q$ possible combinations for the
periodic-$q$ orbit of the $N$-dimensional auxiliary map. The regions where
each tongue exists are given by a system of inequalities that the $N$-tuplet
has to satisfy for the orbit to undergo the right combination of $N$ and
$N-1$ sites in the interface. In Fig.~\ref{blobs-z.ps} we show the tongues
for the velocities $v=1/2,\,2/5,\,1/3,\,2/7$ and $1/4$, computed
numerically, in the minimal $N$-state layers for $N=1,\dots,8$. We call the
tongues in the minimal $N$-state layer the {\it sub-$N$-tongues\/} since
they emanate from the principal mode-locking tongues, the {\it principal
tongues}, for the one-dimensional case. Therefore, from every principal
tongue there are $q$ corresponding sub-$N$-tongues in each $N$-layer. The
structure of these sub-tongues is then self-similar and repeats itself in
every layer as it may be observed in Fig.~\ref{blobs-z.ps}.

\begin{figure}
\centerline{\epsfxsize=8.5cm \epsfbox{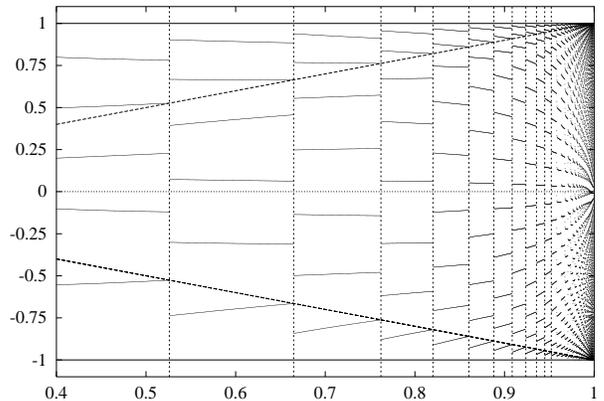}}
\vskip .25cm
\caption{Bifurcation diagram of the attracting
cycle of the interface sites in the $v=1/3$ tongue family. We plot the
stable periodic 3 orbit of the interfacial sites as a function of the
parameter $a$ after 10\,000 transients. The values of $\e$ were taken such
that the pair $(\e,a)$ remained in a path inside the $v=1/3$ mode-locked
tongue (see dashed line inside the tongues in Fig.~\ref{bifu1-3.eps}).}
\label{bifv1_3.ps}\end{figure}

It is interesting to notice that because of the continuity of $v(\e,a)$ all
the sub-$N$-tongues touch each other. Furthermore, the sub-$N$-tongues touch
in a single point, {\it i.e.}~in that particular point the width of the
$\e$-mode-locked plateau (for a fixed value of $a$) is zero. This phenomenon
repeats itself all along each family of sub-$N$-tongues and it happens every
time a site of the interface touches the boundary of the superstable regions
$S_\pm$ when varying the $(\e,a)$-parameters in order to go from one of the
$q$ possible combinations of the interface orbit to the next. In
Fig.~\ref{bifu1-3.eps} we show the state of the lattice at different stages
of the $v=1/3$ tongue. The $\e$-width of the $v=1/3$ tongue is zero in the
transitional case between two different interfacial orbits (cases b), d), f)
and h)). One may consider this switch from one interfacial orbit to the next
as a border-collision bifurcation of the attracting cycle of
the-$N$-dimensional auxiliary map. In order to illustrate when this
bifurcation takes place we show in Fig.~\ref{bifv1_3.ps} the bifurcation
diagram of the attracting cycle of the interface in the $v=1/3$ tongue as
the parameter $a$ is varied. From the figure it is possible to observe that
when a stable point of the interface touches the boundary of the superstable
region (thick dashed diagonal lines $a$ and $-a$) the attracting cycle
changes because a further site is added to it. Where this happens (vertical
dashed lines) the $\e$-width of the tongue is zero (see
Fig.~\ref{bifu1-3.eps}).

\section[Generalizations]{Generalizations}
For every family of tongues for a particular mode-locked ratio there is an
equivalent scenario as the one presented for $v=1/3$. The mode-locking
regions could be computed analytically in every layer along with the
attracting cycles and bifurcation diagrams. This is made possible by the
simple form of the piece-wise linear local map. In more general CMLs, one
way or diffusive, with bistable smooth local maps, the mode-locked regions
still exist though they are not as large as in the piece-wise linear case
because they typically involve an infinite number of sites in the interface
[Carretero, 1997b]. The reduction of the dynamics to a toral map for the
piece-wise linear local map requires a finite number of interfacial sites.
Therefore it seems impossible to apply a toral map reduction of the dynamics
for a generic CML. Nevertheless, if there exists an invariant travelling
interface $h(i-vt)$ it is possible to reduce the dynamics of an infinite
interface to a one-dimensional circle map whose rotation number describes
the velocity of the travelling interface [Carretero {\it et al.}, 1997c].

The border-collision bifurcations corresponding to tran\-si\-tions between
families of sub-tongues prevails for more general CMLs and a similar
scenario as the one depicted here for non-negative gap still exists. On the
other hand, the bor\-der-colli\-sion bi\-fur\-ca\-tions co\-rres\-pon\-ding to 
transitions between sub-tongues of the same family, {\it i.e.}~with the same rotation
number, do not occur for smooth local maps. For the piece-wise linear CML
these bifurcations came from a collision of a site with the border of the
interface. For generic CMLs the interface is the whole interval
$[x_-^*,x_+^*]$ and thus it is impossible to have this kind of  
border-collision bifurcations.

\acknowledgments{
I would like to thank F.~Vivaldi and D.K.~Arrowsmith for fruitful
discussions, comments and corrections. I would also like to thank
A.~Ch\'avez-Ross for continuous support and encouragement. I gratefully
acknowledge DGAPA-UNAM (Mexico) for the financial support during the
preparation of this work.}\bigskip \bigskip

\centerline{\bf REFERENCES}
\bigskip

\noindent Baesens, C., Guckenheimer, J., Kim, S.~\& Mackay, R.~S. [1991]
``Three coupled oscillators: mode-locking, global bifurcations and toroidal
chaos,'' {\it \PD} {\bf 49}, 387--475.
\medskip

\noindent Beck, C. [1994]
``Chaotic cascade model for turbulent velocity distribution,''
{\it \PRE} {\bf 49}(5), 3641--3652.
\medskip

\noindent Carretero-Gonz\'alez, R., Arrowsmith, D.~K.~\& Vivaldi, F. [1997a]
``Mode-locking in \cmls'', {\it \PD} {\bf 103}, 381--403.
\medskip

\noindent Carretero-Gonz\'alez, R. [1997b]
{\it Front propagation and mode-locking in \cmls}.
PhD thesis, Queen Mary and Westfield College, London, U.K.
\medskip

\noindent Carretero-Gonz\'alez, R., Arrowsmith, D.~K.~\& Vivaldi, F. [1997c]
``Reduction dynamics for travelling fronts in {\cmls},'' in preparation.
\medskip

\noindent Chat\'e, H.~\& Manneville, P. [1989]
``Coupled map lattices as cellular automata,''
{\it \JSP} {\bf 56}(3/4), 357--370.
\medskip

\noindent Chat\'e, H.~\& Manneville, P. [1990]
``Using {\cmls} to unveil structures in the space of cellular automata,''
{\it Springer proceedings in physics, Cellular automata and
modeling of complex physical systems} {\bf 46}, 298--309.
\medskip

\noindent Hassell, M.~P., Miramontes, O., Rohani, P.~\& May, R.~M. [1995]
``Appropriate formulations for dispersal in spatially structured models,''
{\it \JAE} {\bf 64}, 662--664.
\medskip

\noindent Kaneko, K. [1983]
``Transition from torus to chaos accompanied by frequency lockings
with symmetry breaking,'' {\it \PTP} {\bf 69}(5), 1427.
\medskip

\noindent Kaneko, K. [1984]
``Period-doubling of kink-antikink patterns, quasiperiodicity in
anti-ferro-like structures and spatial intermittency in coupled logistic
lattice,'' {\it \PTP} {\bf 72}(3), 480--486.
\medskip

\noindent Kaneko, K. [1985]
``Spatial Period-doubling in open flow,'' {\it \PL} {\bf 111}(7), 321--325.
\medskip

\noindent Kaneko, K. [1986]
``Turbulence in \cmls,'' {\it \PD} {\bf 18}, 475--476.
\medskip

\noindent Kaneko, K. [1989]
``Spatiotemporal chaos in one- and two-dimensional \cmls,''
{\it \PD} {\bf 37}, 60--82.
\medskip

\noindent Kaneko, K. [1992]
``Global travelling wave triggered by local phase slips,''
{\it \PRL} {\bf 69}(6), 905--908.
\medskip

\noindent Kaneko, K. [1993]
``Chaotic travelling waves in a \cml,'' {\it \PD} {\bf 68}, 299--317.
\medskip

\noindent Maistrenko, Y.~L., Maistrenko, V.~L., Vikul, S.~I.~\& Chua, L.~O. [1995]
``Bifurcations of attracting cycles from the delayed Chua's
circuit,'' {\it \IJBC} {\bf 5}(3), 653--671.
\medskip

\noindent Sol\'e, R.~V.~\& Bascompte, J. [1995]
``Measuring chaos from spatial information,'' {\it \JTB} {\bf 175}, 139--147.
\medskip

\noindent Willeboordse F.~H.~\& Kaneko, K. [1995]
``Pattern dynamics of a {\cml} for open flow,'' {\it \PD} 101--128.
\medskip

\noindent Yanagita, T.~\& Kaneko, K. [1993]
``{\Cml} model for convection,'' {\it \PLA} {\bf 175}, 415--420.
\medskip

\vfill
\eject
\end{document}